\documentclass[a4paper, 12pt]{article} 
\usepackage[right=1.1in,left=1.1in,top=1.1in,bottom=1.1in]{geometry}
\usepackage{graphicx} 
\usepackage{braket}
\usepackage{mathrsfs}
\usepackage[T1]{fontenc} 
\usepackage{amsmath}
\usepackage{pxfonts}
\usepackage[T1]{fontenc}
\usepackage{amssymb}
\usepackage{natbib}
\usepackage{epstopdf}
\usepackage{setspace}
\usepackage{enumitem}
\usepackage{sectsty}
\usepackage{wrapfig} 
\sectionfont{\large}
\bibpunct{(}{)}{,}{a}{}{,}
\usepackage{tikz}   
\usepackage{tikz-3dplot} 

\newcommand{\qed}{\nobreak \ifvmode \relax \else
      \ifdim\lastskip<1.5em \hskip-\lastskip
      \hskip1.5em plus0em minus0.5em \fi \nobreak
      \vrule height0.75em width0.5em depth0.25em\fi}

\usepackage{bbm}
\usepackage{MnSymbol}
\usepackage[all]{xy}

\usepackage{xcolor,colortbl,array,amssymb}

   \usepackage{braket}
   \usepackage{amssymb}
\usepackage{multicol}

\makeatletter

\title{ \Large{Time's Arrow in a Quantum Universe:  }
\\ \Large{On the Status of Statistical Mechanical Probabilities}
} 

\author{Eddy Keming Chen\thanks{Department of Philosophy, 106 Somerset Street, Rutgers University, New Brunswick, NJ 08901, USA. Website: www.eddykemingchen.net. Email: eddy.chen@rutgers.edu  }}  

\date{\small{Forthcoming in Valia Allori (ed.), \emph{Statistical Mechanics and Scientific Explanation: Determinism, Indeterminism and Laws of Nature},  World Scientific} } 

\begin{document}
\bibliographystyle{apalike}

\maketitle 



\begin{abstract}

In a quantum universe with a strong arrow of time, it is standard to postulate that the initial wave function started in a particular macrostate---the special low-entropy macrostate selected by the Past Hypothesis. Moreover, there is an additional postulate about statistical mechanical probabilities according to which the initial wave function is a ``typical'' choice in the macrostate (the Statistical Postulate). Together, they support a probabilistic version of the Second Law of Thermodynamics:  typical initial wave functions will increase in entropy. Hence, there are two sources of randomness in such a universe: the quantum-mechanical probabilities of the Born rule and the statistical mechanical probabilities of the Statistical Postulate. 

I propose a new way to understand time's arrow in a quantum universe. It is based on what I call the Thermodynamic Theories of Quantum Mechanics. According to this perspective, there is a \emph{natural} choice for the initial quantum state of the universe, which is given  by not a wave function but by a density matrix. The density matrix plays a microscopic role: it appears in the fundamental dynamical equations of those theories. The density matrix also plays a macroscopic / thermodynamic role: it is exactly the (normalized) projection operator onto the Past Hypothesis subspace (of the  Hilbert space of the universe). Thus, given an initial subspace, we obtain a unique choice of the initial density matrix. I call this property \emph{the conditional uniqueness} of the initial quantum state. The conditional uniqueness provides a new and general strategy to eliminate statistical mechanical probabilities  in the fundamental physical theories, by which we can reduce the two sources of randomness to only the quantum mechanical one. I also explore the idea of an \emph{absolutely unique} initial quantum state, in a way that might realize Penrose's idea (1989) of a strongly deterministic universe.

\end{abstract}

\hspace*{3,6mm}\textit{Keywords: time's arrow, Past Hypothesis, Statistical Postulate, the Mentaculus Vision, typicality, unification, foundations of probability, quantum statistical mechanics, wave function realism, quantum ontology, density matrix, initial condition of the universe, strong determinism}   


\begingroup
\singlespacing
\tableofcontents
\endgroup

\vspace{20pt} 



\nocite{ AlbertLPT, loewer2004david, lebowitz2008time, goldstein2001boltzmann, durr1992quantum, durr2012quantum, goldstein2013reality, goldstein2010approachB, goldstein2010approach, goldstein2010normal, durr2005role, bell1980broglie, goldstein2012typicality, ney2013wave, ChenOurFund, ChenHSWF, chen2017intrinsic, allori2013primitive, allori2008common, loewer2016mentaculus, LewisPP2, sep-time-thermo, north2011time, coen2010serious}

\section{Introduction}

In a quantum universe with a strong arrow of time (large entropy gradient), it is standard to attribute the temporal asymmetry to a special boundary condition. This boundary condition is a macrostate of extremely low entropy. David Albert (2000)  calls it the Past Hypothesis. The quantum version of the Past Hypothesis dictates that the initial wave function of the universe lies within the low-entropy macrostate. Mathematically, the macrostate is represented by a low-dimensional subspace in the total Hilbert space, which corresponds to low Boltzmann entropy. 

The Past Hypothesis is accompanied by another postulate about statistical mechanical probabilities according to which the initial wave function is a ``typical'' choice in the macrostate (the Statistical Postulate). Together, the Past Hypothesis and the Statistical Postulate support a probabilistic version of the Second Law of Thermodynamics:  typical initial wave functions will increase in entropy. 

The standard theory, which has its origin in Boltzmann's statistical mechanics and has been substantially developed in the last century, is a simple and elegant way of understanding time's arrow in a quantum universe. It has two features:
\begin{enumerate}
\item Although the theory restricts the choices of initial quantum states of the universe, it does not select a unique one.
\item The theory postulates statistical mechanical probabilities (or  a typicality measure) on the fundamental level. They are  \emph{prima facie} on a par with and in addition to the quantum mechanical probabilities. 
\end{enumerate}
Hence, there are two sources of randomness in such a universe: the quantum-mechanical probabilities in the Born rule and the statistical mechanical probabilities in the Statistical Postulate.  It is an interesting conceptual question how we should understand the two kinds of probabilities in the standard theory. 

In this paper, I propose a new way to understand time's arrow in a quantum universe. It is based on what I call the Thermodynamic Theories of Quantum Mechanics (TQM). According to this perspective, there is a \emph{natural} choice for the initial quantum state of the universe, which is given not by a wave function but by a density matrix. Moreover, the density matrix enters into the fundamental equations of quantum mechanics. Hence, it plays a microscopic role. Furthermore, the density matrix is exactly the (normalized) projection operator onto the Past Hypothesis macrostate, represented by a subspace in the Hilbert space of the universe. 

In \cite{chen2018IPH}, I introduced the Initial Projection Hypothesis (IPH) as a new postulate about a natural initial density matrix of the universe. In that paper, I focused on its relevance to the debate about the nature of the quantum state. In \cite{chen2019quantum1}, I focused on its empirical equivalence with the Past Hypothesis. In this paper, I focus on its relevance to the status of statistical mechanical probabilities. According to the Initial Projection Hypothesis, given a choice of an initial subspace, we have a unique choice of the initial density matrix. This property is called \emph{conditional uniqueness}, which will be sufficient to eliminate statistical mechanical probabilities at the level of fundamental physics. Thus, in contrast to the standard theory,  TQM has the following two features:
\begin{enumerate}
\item The theory selects a unique initial quantum state of the universe, given a choice of the Past Hypothesis subspace. [Conditional Uniqueness]
\item Because of conditional uniqueness, the theory does not postulate statistical mechanical probabilities (or  a typicality measure) at the level of fundamental physics.    
\end{enumerate}
That is, TQM provides a new and general strategy for eliminating statistical mechanical probabilities at the fundamental level. To be sure, statistical mechanical probabilities can still emerge as useful tools of analysis. TQM satisfies conditional uniqueness, which is sufficient for the purpose at hand. But we might also wonder whether TQM may lead to absolute uniqueness, in the sense that there is a unique initial quantum state of the universe that does not depend on a prior selection of a unique initial subspace. Such absolute uniqueness will go some way towards realizing a strongly deterministic universe in the sense of \cite{roger1989emperor}. 

Here is the roadmap of the paper. In \S2, I will review the standard (Boltzmannian) way of understanding temporal asymmetry in classical and quantum mechanics. In \S3, I introduce the Thermodynamic Theories of Quantum Mechanics. First, I  formulate the Initial Projection Hypothesis, which is the key component of TQM and an alternative to the Past Hypothesis. Next, I  focus on the property of conditional uniqueness, the elimination of statistical mechanical probabilities at the fundamental level, the emergence of such probabilities as useful tools of analysis, and the possibility of absolute uniqueness and strong determinism. I also discuss  the (lack of) analogues in classical statistical mechanics,  implications for the status of the quantum state, Lorentz invariance, theoretical unity, generalizations to other cosmological initial conditions, and  relation to other proposals in the literature. 

\section{Statistical Mechanics and Time's Arrow}

Statistical mechanics concerns macroscopic systems such as a gas in a box. It is an important subject for  understanding the arrow of time, since thermodynamic systems are irreversible. A gas in a box can be described as a system of $N$ particles, with $N> 10^{20}$. If the system is governed by classical mechanics, although it is difficult to solve the equations exactly, we can still use classical statistical mechanics (CSM) to describe its statistical behaviors, such as approach to thermal equilibrium suggested by the Second Law of Thermodynamics. Similarly, if the system is governed by quantum mechanics, we can use quantum statistical mechanics (QSM) to describe its statistical behaviors. Generally speaking, there are two different views on CSM: the individualistic view and the ensemblist view. For concreteness, we will adopt the individualistic view in this paper, and we will start with a review of CSM, which should be more familiar to researchers in the foundations of physics. 

\subsection{Classical Statistical Mechanics}

Let us review the basic elements of CSM on the individualistic view.\footnote{Here I follow the discussion in \cite{goldstein2010approachB}. \S 2.1 and \S 2.2 do not intend to be rigorous axiomatizations of  CSM and QSM. They are only summaries of the key concepts that are important for appreciating the main ideas in the later sections.} For concreteness, let us consider a classical-mechanical system with $N$ particles in a box $\Lambda = [0, L]^3 \subset \mathbb{R}^3$ and a Hamiltonian $H = H(X) = H(\boldsymbol{q_1}, ..., \boldsymbol{q_N}; \boldsymbol{p_1},...,\boldsymbol{p_n})$.

\begin{enumerate}
\item Microstate: at any time $t$, the microstate of the system is given by a point in a $6N$-dimensional phase space,
\begin{equation}
X=(\boldsymbol{q_1}, ..., \boldsymbol{q_N}; \boldsymbol{p_1},...,\boldsymbol{p_n}) \in \Gamma_{total} \subseteq \mathbb{R}^{6N},
\end{equation}
where $\Gamma_{total}$ is the total phase space of the system. 

\item Dynamics: the time dependence of $X_t =(\boldsymbol{q_1}(t), ..., \boldsymbol{q_N}(t); \boldsymbol{p_1}(t),...,\boldsymbol{p_n}(t))$ is given by the Hamiltonian equations of motion:
\begin{equation}\label{HE}
\frac{d \boldsymbol{q_i}}{d t} = \frac{\partial H}{\partial \boldsymbol{p_i}} \text{  ,  } \frac{d \boldsymbol{p_i}}{d t} = - \frac{\partial H}{\partial \boldsymbol{q_i}}.
\end{equation}

\item Energy shell: the physically relevant part of the total phase space is the energy shell $\Gamma \subseteq \Gamma_{total}$ defined as:
\begin{equation}
\Gamma = \{X\in \Gamma_{total}: E\leq H(x) \leq E+\delta E\}.
\end{equation}

We only consider microstates in $\Gamma.$

\item Measure: the measure $\mu_V$ is the standard Lebesgue measure on phase space, which is the volume measure on $\mathbb{R}^{6N}$. 

\item Macrostate: with a choice of macro-variables, the energy shell $\Gamma$ can be partitioned into macrostates $\Gamma_{\nu}$: 
\begin{equation}
\Gamma = \bigcup_{\nu} \Gamma_{\nu}.
\end{equation}
A macrostate is composed of microstates that share similar macroscopic features (similar values of the macro-variables), such as volume, density, and pressure. 

\item Unique correspondence: every phase point $X$ belongs to one and only one $\Gamma_{\nu}$. (This point is implied by (5). But we make it explicit to better contrast it with the situation in QSM.)

\item Thermal equilibrium: typically, there is a dominant macrostate $\Gamma_{eq}$ that has almost the entire  volume with respect to $\mu_V$:
\begin{equation}
\frac{\mu_V(\Gamma_{eq})}{\mu_V(\Gamma)} \approx 1.
\end{equation}

A system is in thermal equilibrium if its phase point $X\in \Gamma_{eq}.$ 

\item Boltzmann Entropy: the Boltzmann entropy of a classical-mechanical system in microstate $X$ is given by:
\begin{equation}
S_B (X) = k_B \text{log} (\mu_V(\Gamma(X))),
\end{equation}
where $\Gamma(X)$ denotes the macrostate containing $X$. The thermal equilibrium state thus has the maximum entropy. 

\item Low-Entropy Initial Condition: when we consider the universe as a classical-mechanical system, we postulate a special low-entropy boundary condition, which David Albert calls \emph{the Past Hypothesis}: 
\begin{equation}
 X_{t_0} \in \Gamma_{PH} \text{ , } \mu_V(\Gamma_{PH}) \ll \mu_V(\Gamma_{eq}) \approx \mu_V(\Gamma),
\end{equation}
where $\Gamma_{PH}$ is the Past Hypothesis macrostate with volume much smaller than that of the equilibrium macrostate.  Hence,  $S_B (X_{t_0})$, the Boltzmann entropy of the  microstate at the boundary, is very small compared to that of thermal equilibrium. 

\item A central task of CSM is to establish mathematical results that demonstrate (or suggest) that  $\mu_V-$most microstates  will approach thermal equilibrium. 
\end{enumerate}

\subsection{Quantum Statistical Mechanics}

Let us now turn to quantum statistical mechanics (QSM). For concreteness, let us consider a quantum-mechanical system with $N$ fermions (with $N> 10^{20}$) in a box $\Lambda = [0, L]^3 \subset \mathbb{R}^3$ and a Hamiltonian $\hat{H}$. For concreteness, I will focus on the individualistic view of QSM. The main difference with CSM is that they employ different state spaces. The classical state space is the phase space while the quantum state space is the Hilbert space. However, CSM and QSM are conceptually similar, as we can see from below.\footnote{Here I follow the discussions in \cite{goldstein2010approach} and \cite{goldstein2010approachB}.} 

\begin{figure}[h]
\centerline{\includegraphics[scale=0.46]{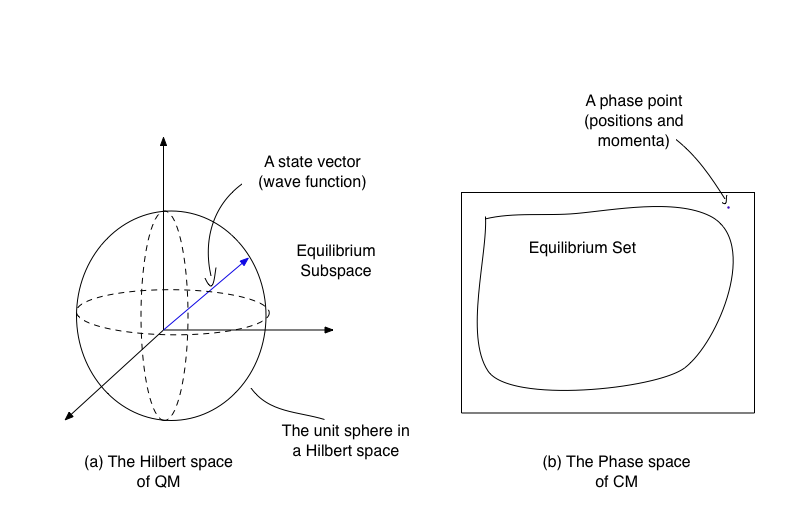}}
\caption{QM and CM employ different state spaces.  These pictures are not drawn to scale, and they are drawn with much fewer dimensions. In (a), we only draw a three-dimensional equilibrium subspace, which is embedded in a higher-dimensional energy shell. In (b), we only draw a two-dimensional energy surface, which includes a large equilibrium set.}
\end{figure}

\begin{enumerate}
\item Microstate: at any time $t$, the microstate of the system is given by a normalized (and anti-symmetrized) wave function:
\begin{equation}
\psi(\boldsymbol{q_1}, ..., \boldsymbol{q_N}) \in \mathscr{H}_{total} = L^2 (\Lambda^{N}, \mathbb{C}^k) \text{ , } \parallel \psi \parallel_{L^2} = 1,
\end{equation}
where $\mathscr{H}_{total} = L^2 (\Lambda^{N},  \mathbb{C}^k)$ is the total Hilbert space of the system. 

\item Dynamics: the time dependence of $\psi(\boldsymbol{q_1}, ..., \boldsymbol{q_N}; t)$ is given by the Schr\"odinger equation: 
\begin{equation}\label{SE}
 i\hbar \frac{\partial \psi}{\partial t} = \hat{H} \psi.
\end{equation}

\item Energy shell: the physically relevant part of the total Hilbert space is the subspace (``the energy shell''):
\begin{equation}
\mathscr{H} \subseteq \mathscr{H}_{total} \text{ , } \mathscr{H} = \text{span} \{ \phi_\alpha : E_\alpha \in [E, E+\delta E ]  \},
\end{equation}
This is the subspace (of the total Hilbert space) spanned by energy eigenstates $\phi_\alpha$ whose eigenvalues $E_\alpha$ belong to the $[E, E+\delta E]$ range.  Let $D = \text{dim} \mathscr{H}$, the number of energy levels between $E$ and $E+\delta E$. 

We only consider wave functions $\psi$ in $\mathscr{H}$.

\item Measure: given a subspace $\mathscr{H}$, the measure $\mu_S$ is  the surface area measure on the unit sphere in that subspace $\mathscr{S}(\mathscr{H})$.\footnote{For simplicity, here we assume that the subspaces we deal with are finite-dimensional. In cases where the Hilbert space is infinite-dimensional, it is an open and challenging technical question.  For example, we could use Gaussian measures in infinite-dimensional spaces, but we  no longer  have uniform probability distributions. }

\item Macrostate: with a choice of macro-variables,\footnote{For technical reasons,  \cite{von1955mathematical}  suggests that we round up these macro-variables (represented by quantum observables) so as to make the  observables commute. See \cite{goldstein2010long} \S2.2 for a discussion of von Neumann's ideas. } the energy shell $\mathscr{H}$ can be orthogonally decomposed into macro-spaces:
\begin{equation}
\mathscr{H} = \oplus_\nu \mathscr{H}_\nu \text{ , } \sum_\nu \text{dim}\mathscr{H}_\nu  = D
\end{equation}
Each $\mathscr{H}_\nu$ corresponds more or less to small ranges of values of macro-variables that we have chosen in advance. 

\item Non-unique correspondence: typically, a wave function is in a superposition of macrostates and is not entirely in any one of the macrospaces. However, we can make sense of situations where $\psi$ is (in the Hilbert space norm) very close to a macrostate $\mathscr{H}_\nu$: 
\begin{equation}\label{close}
\bra{\psi} P_{\nu}  \ket{\psi} \approx 1,
\end{equation}
where $P_{\nu}$ is the projection operator into $\mathscr{H}_{\nu}$. This means that  $\ket{\psi}$ lies almost entirely in $\mathscr{H}_{\nu}$. (This is different from CSM, where every phase point lies entirely within some macrostate.)

\item Thermal equilibrium: typically, there is a dominant macro-space $\mathscr{H}_{eq}$ that has a dimension that  is almost equal to D: 
\begin{equation}
\frac{\text{dim} \mathscr{H}_{eq}}{\text{dim} \mathscr{H}} \approx 1.
\end{equation}
A system with wave function $\psi$ is in equilibrium if the wave function $\psi $ is very close to $\mathscr{H}_{eq}$ in the sense of (\ref{close}):  $\bra{\psi} P_{eq}  \ket{\psi} \approx 1.$

\emph{Simple Example.} Consider a gas consisting of $n = 10^{23}$ atoms in a box $\Lambda \subseteq \mathbb{R}^3.$ The system is governed by quantum mechanics. We orthogonally decompose the Hilbert space $\mathscr{H}$ into 51 macro-spaces: $\mathscr{H}_0 \oplus \mathscr{H}_2  \oplus \mathscr{H}_4  \oplus...  \oplus \mathscr{H}_{100},$ where $\mathscr{H}_\nu$ is the subspace corresponding to the macrostate such that the number of atoms in the left half of the box is between $(\nu-1) \%$ and $(\nu+1) \%$ of $n$, with the endpoints being the exceptions: $\mathscr{H}_0$ is the interval of $0\% - 1\%$ and $\mathscr{H}_{100}$ is the interval of $99 \%- 100\%$. In this example, $\mathscr{H}_{50}$ has the overwhelming majority of dimensions and is thus the equilibrium macro-space. A system whose wave function is very close to $\mathscr{H}_{50}$ is in equilibrium (for this choice of macrostates).

\item Boltzmann Entropy: the Boltzmann entropy of a quantum-mechanical system with wave function $\psi$ that is very close to a macrostate $\nu$ is given by:
\begin{equation}\label{Boltzmann}
S_B (\psi) = k_B \text{log} (\text{dim} \mathscr{H}_\nu ),
\end{equation}
where $\mathscr{H}_\nu$ denotes the subspace containing almost all of $\psi$ in the sense of (\ref{close}). The thermal equilibrium state thus has the maximum entropy: 
\begin{equation}
S_B (eq) = k_B \text{log} (\text{dim} \mathscr{H}_{eq} ) \approx  k_B \text{log} (D),
\end{equation}
where \emph{eq} denotes the equilibrium macrostate. 

\item Low-Entropy Initial Condition: when we consider the universe as a quantum-mechanical system, we postulate a special low-entropy boundary condition on the universal wave function---the quantum-mechanical version of \emph{the Past Hypothesis}: 
\begin{equation}
\Psi(t_0) \in \mathscr{H}_{PH} \text{ , } \text{dim} \mathscr{H}_{PH} \ll \text{dim}\mathscr{H}_{eq} \approx \text{dim} \mathscr{H}
\end{equation}
where $\mathscr{H}_{PH}$ is the Past Hypothesis macro-space with dimension much smaller than that of the equilibrium macro-space.\footnote{Again, we  assume that $\mathscr{H}_{PH}$ is finite-dimensional, in which case we can use the surface area measure on the unit sphere as the typicality measure for \# 10. It remains an open question in QSM about how to formulate the low-entropy initial condition when the initial macro-space is infinite-dimensional.} Hence, the initial state has very low entropy in the sense of (\ref{Boltzmann}). 

\item A central task of QSM is to establish mathematical results that demonstrate (or suggest) that $\mu_S-$most (maybe even all) wave functions of small subsystems, such as gas in a box,  will approach thermal equilibrium. 
\end{enumerate}


If the microstate $\psi$ of a system is close to some macro-space $\mathscr{H}_{\nu}$ in the sense of (\ref{close}), we can say that the macrostate of the system is $\mathscr{H}_{\nu}$. The macrostate $\mathscr{H}_{\nu}$ is naturally associated with a density matrix: 
\begin{equation}\label{ID}
\hat{W}_{\nu} = \frac{I_{\nu}}{dim \mathscr{H}_{\nu}},
\end{equation}
where $I_{\nu}$ is the projection operator onto  $\mathscr{H}_{\nu}$. The density matrix also obeys a dynamical equation; it evolves according to the von Neumann equation:
\begin{equation}\label{VNM}
i \hbar \frac{d \hat{W}(t)}{d t} = [\hat{H},  \hat{W}].
\end{equation}

Given the canonical correspondence between a subspace and its normalized projection operator, $\hat{W}_{\nu}$ is also a representation of the subspace $\mathscr{H}_{\nu}$. It can be decomposed into wave functions, but the decomposition is not unique. Different measures can give rise to the same density matrix.  One such choice is $\mu_S(d\psi)$, the uniform distribution over wave functions: 
\begin{equation}\label{MacroW}
\hat{W}_{\nu} = \int_{\mathscr{S}(\mathscr{H_{\nu}})} \mu_S(d\psi) \ket{\psi} \bra{\psi}.
\end{equation}
In (\ref{MacroW}), $\hat{W}_{\nu}$ is defined with a choice of measure on wave functions in $\mathscr{H_{\nu}}$. However, we should not be misled into thinking that the density matrix is derivative of  wave functions. What is  intrinsic to a density matrix is its geometrical meaning in the Hilbert space. In the case of $\hat{W}_{\nu}$,  as shown in the canonical description (\ref{ID}), it is just a normalized projection operator.

\subsection{$\Psi_{PH}$-Quantum Theories}

\begin{figure}[h]
\centerline{\includegraphics[scale=0.46]{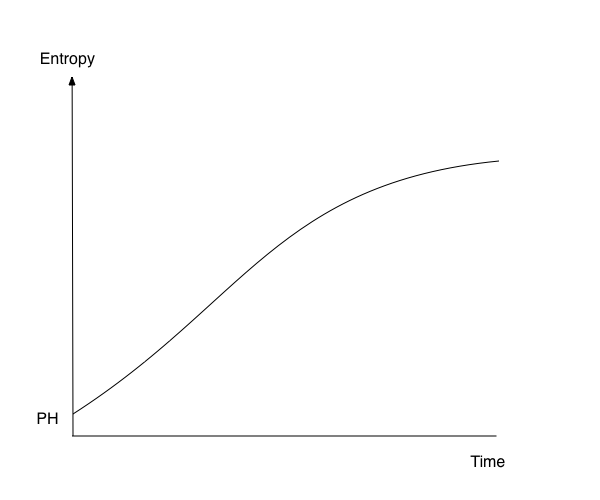}}
\end{figure}

If we treat the universe as a quantum system, then we can distill a  picture of the fundamental physical reality from the standard Boltzmannian QSM (individualistic perspective). The universe is described by a quantum state represented by a universal wave function. It starts in a Past Hypothesis macrostate (\S2.2 \#9) and is selected randomly according to the probability  specified by the Statistical Postulate  (\S2.2 \#4). It evolves by the quantum dynamics---the Schr\"odinger equation (\S2.2 \#2). That is, it has three fundamental postulates (fundamental lawlike statements\footnote{Barry Loewer calls the joint system---the package of laws that includes PH and SP in addition to the dynamical laws of physics---the Mentaculus Vision. For developments and defenses of the nomological account of the Past Hypothesis and the Statistical Postulate, see \cite{albert2000time, LoewerCatSLaw, wallace2011logic, wallace2012emergent} and \cite{loewer2016mentaculus}. Albert and Loewer are writing mainly in the context of CSM. The Mentaculus Vision is supposed to provide a ``probability map of the world.'' As such, it requires one to take the probability distribution very seriously.}): 
\begin{enumerate}
\item The Schr\"odinger equation.
\item The Past Hypothesis. 
\item The Statistical Postulate. 
\end{enumerate}
However, this theory by itself faces the measurement problem. To solve the measurement problem, we can combine it with three well-known strategies: 
\begin{itemize}
\item Bohmian mechanics (BM): the fundamental ontology in addition to the quantum state also includes point particles with precise locations, and the fundamental dynamics also includes a guidance equation that relates the wave function to the velocity of the particles. 
\item Everettian mechanics (S0): the fundamental ontology consists in just the quantum state evolving unitarily according to the Schr\"odinger equation.
\item GRW spontaneous collapse theory (GRW0): the fundamental ontology consists in just the quantum state evolving by the Schr\"odinger equation, but the unitary evolution is interrupted by a spontaneous collapse mechanism. 
\end{itemize}
The universal wave function $\Psi$ is central to standard formulations of the above theories. The Past Hypothesis and the Statistical Postulate need to be added to them to account for the arrow of time. Let us label them $\Psi_{PH}$-BM, $\Psi_{PH}$-S0, and $\Psi_{PH}$-GRW0. They are all instances of what I call  $\Psi_{PH}$-quantum theories.

There are also Everettian and GRW theories with additional fundamental ontologies in physical space-time, such as Everettian theory with a mass-density ontology (S0), GRW theory with a mass-density ontology (GRWm), and GRW theory with a flash ontology (GRWf). See \cite{allori2008common, allori2010many} for discussions. Thus, we  also have $\Psi_{PH}$-Sm,  $\Psi_{PH}$-GRWm, and $\Psi_{PH}$-GRWf.

\section{The Thermodynamic Theories of Quantum Mechanics}

The  $\Psi_{PH}$-quantum theories attempt to solve the measurement problem and account for the arrow of time.  However, each $\Psi_{PH}$-quantum theory admits many choices for the initial quantum state.  This is because the Past Hypothesis subspace $\mathscr{H}_{PH}$, although being small compared to the full energy shell, is still compatible with many wave functions. One might wonder whether there is some hypothesis  that could determine a unique initial quantum state of the universe. There is an interesting consequence if that can be done:  it may bring us closer to what \cite{roger1989emperor} calls \emph{strong determinism}, the idea that the entire history of the universe is fixed by the theory, with laws that specify both a deterministic dynamics and a (nomologically) unique initial state of the universe (see \S3.5). As we shall see, there are other benefits of having a unique initial quantum state of the universe, such as the elimination of statistical mechanical probabilities (\S3.2.1). 

Is it possible to  narrow down the initial choices to exactly one? I believe that we can, but we need to make some conceptual changes to the standard theory. Our proposal consists in three steps: 
\begin{enumerate}
\item Allowing the universe to be in a fundamental mixed state represented by a density matrix. 
\item Letting the density matrix  enter into the fundamental dynamical equations. 
\item Choosing a natural density matrix associated with the Past Hypothesis subspace $\mathscr{H}_{PH}$.
\end{enumerate}
The natural density matrix together with its fundamental dynamical equations  define a new class of quantum theories, which we will call the \emph{Thermodynamic Theories of Quantum Mechanics}. As we explain below, all of them will  have a conditionally unique initial quantum state. 

\subsection{The Initial Projection Hypothesis}

The standard approach to the foundations of quantum mechanics assumes that the universe is described by a \emph{pure} quantum state $\Psi(t)$, which is represented as a vector in the (energy shell of the) Hilbert space of the system. However, it is also possible to describe the universe by a \emph{mixed} quantum state $W(t)$, which has its own geometric meaning in the Hilbert space. 

On the latter perspective, the quantum state of the universe is in a mixed state $W$, which becomes  the central object of  quantum theory. $W$ obeys its own unitary dynamics---the von Neumann equation (\ref{VNM}), which is a generalization of the Schr\"odinger equation (\ref{SE}). Moreover, we can write down density-matrix versions of Bohmian mechanics, Everettian mechanics, and GRW spontaneous collapse theory. (See Appendix for the mathematical details.) Given these equations, the density matrix of the universe plays a microscopic role: it enters into the fundamental dynamical equations of those theories. As a result, it guides Bohmian particles, gives rise to GRW collapses, and realizes the emergent multiverse of Everett. I call this view \emph{Density Matrix Realism}, in contrast with Wave Function Realism, the view that the central object of quantum theory is the universal wave function.\footnote{Density Matrix Realism may be unfamiliar to some people, as we are used to take the mixed states to represent our \emph{epistemic uncertainties of the actual pure state} (a wave function).  The proposal here is that the density matrix directly represents the actual quantum state of the universe; there is no further fact about which is the actual wave function. In this sense, the density matrix is ``fundamental.'' In fact,  this idea has come up  in the foundations of physics. See, for example, \cite{durr2005role, maroney2005density}, \cite{wallace2011logic, wallace2012emergent},  and \cite{wallace2016probability}. I discuss  this idea in more detail in \cite{chen2018IPH} \S3}.

According Density Matrix Realism, the initial quantum state is also represented by a density matrix. Now, we know that to account for the arrow of time we need to postulate a low-entropy initial condition---the Past Hypothesis. Moreover, under Wave Function Realism, it is also necessary to postulate a Statistical Postulate. 

Now, we make a crucial observation. If the initial quantum state is a density matrix, then there is a natural choice for it given the Past Hypothesis subspace. The natural choice is the normalized projection operator onto  $\mathscr{H}_{PH}$:
\begin{equation}\label{PHID}
\hat{W}_{IPH} (t_0) = \frac{I_{PH}}{dim \mathscr{H}_{PH}},
\end{equation}
where $t_0$ represents a temporal boundary of the universe, $I_{PH}$ is the projection operator onto   $\mathscr{H}_{PH}$, $dim$ counts the dimension of the Hilbert space, and $dim\mathscr{H}_{PH} \ll dim\mathscr{H}_{eq}$. Since the quantum state at $t_0$ has the lowest entropy, we call $t_0$ the initial time.  In  \cite{chen2018IPH}, I call (\ref{PHID}) the \emph{Initial Projection Hypothesis} (IPH). In words: the initial density matrix of the universe is the normalized projection onto the PH-subspace.

\begin{figure}[h]
\centerline{\includegraphics[scale=0.46]{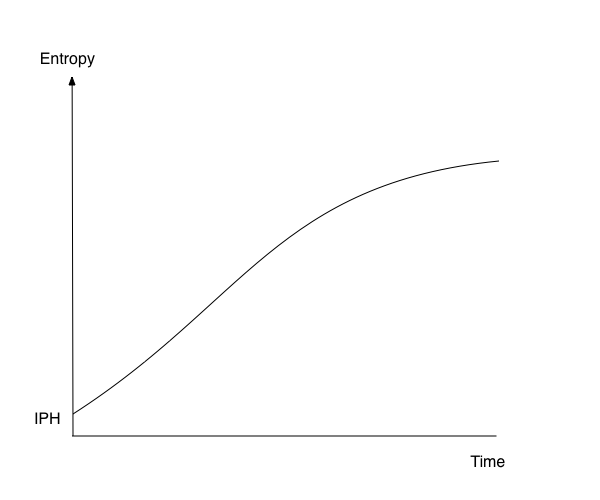}}
\end{figure}

The projection operator onto $\mathscr{H}_{PH}$ contains no more and no less information than what is contained in the subspace itself. There is a natural correspondence between a subspace and its projection operator. If we specify the subspace, we know what its projection operator is, and vice versa. Since the projection operator onto a subspace carries no more information than that subspace itself, the projection operator is no more complex than $\mathscr{H}_{PH}$. This is different from $\Psi_{PH}$, which is confined by PH to be a vector inside $\mathscr{H}_{PH}$. A vector carries more information than the subspace it belongs to, as specifying a subspace is not sufficient to determine  a vector. (For example, to determine a vector in an 18-dimensional subspace of a 36-dimensional vector space, we need 18 coordinates in addition to specifying the subspace. The higher the dimension of the subspace, the more information is needed to specify the vector.) If PH had fixed  $\Psi_{PH}$ (the QSM microstate), it would have required much more information and  become a much more complex posit. PH as it is  determines $\Psi_{PH}$ only up to an equivalence class (the QSM macrostate). 

 I propose, then, that we replace the Past Hypothesis with the Initial Projection Hypothesis, the Schr\"odinger equation with the von Neumann equation, and the universal wave function with a universal density matrix.  In contrast to $\Psi_{PH}$ theories, there are only two corresponding  fundamental postulates (fundamental laws of physics) in the new theory: 
\begin{enumerate}
\item The von Neumann equation.
\item The Initial Projection Hypothesis. 
\end{enumerate}
Notice that IPH defines a unique initial quantum state, given a choice of the initial PH subspace. There is only one state that is possible, not a collection of states, in the PH subspace.  We call this \emph{conditional uniqueness}. The quantum state $\hat{W}_{IPH} (t_0)$ is informationally equivalent to the constraint that PH imposes on the initial microstates. Given the PH subspace, $\hat{W}_{IPH} (t_0)$ is singled out by the data in PH (we will come back to this point in \S3.2). Consequently, on the universal scale, we do not need to impose an additional probability or typicality measure on the Hilbert space. $\hat{W}_{IPH} (t_0)$ is mathematically equivalent to an integral over projection onto each normalized state vectors (wave functions) compatible with PH \emph{with respect to $\mu_S(d\psi)$}. Of course, we are not defining $\hat{W}_{IPH} (t_0)$ in terms of state vectors. Rather, we are thinking of $\hat{W}_{IPH} (t_0)$ as a geometric object in the Hilbert space:  the (normalized) projection operator onto $\mathscr{H}_{PH}$.  That is the \emph{intrinsic} understanding of the density matrix. (In \S3.2 and \S3.3 we will discuss different versions of PH and their relevance to conditional uniqueness. In \S3.5 we will discuss why this strategy does not work so well in CSM.) 

As before, the above theory by itself faces the measurement problem. To solve the measurement problem, we can combine it with the well-known strategies of BM, EQM, and GRW. We will label them as W$_{IPH}$-BM, W$_{IPH}$-EQM, and W$_{IPH}$-GRW.  (See the Appendix for mathematical details.) Together, we  call them  \emph{Density-Matrix Quantum Theories with the Initial Projection Hypothesis} (W$_{IPH}$-QM).   Moreover, since they are motivated by considerations of the thermodynamic macrostate of the early universe, we  also call them the \emph{Thermodynamic Theories of Quantum Mechanics} (TQM).  

In W$_{IPH}$-QM, the density matrix takes on a central  role as the quantum microstate. Besides the low-entropy initial condition, it is also necessary to reformulate some definitions in quantum statistical mechanics:
\begin{itemize}
\item[6'] Non-unique correspondence: typically, a density matrix is in a superposition of macrostates and is not entirely in any one of the macrospaces. However, we can make sense of situations where $W$ is very close to a macrostate $\mathscr{H}_\nu$: 
\begin{equation}\label{Wclose}
\text{tr} (W P_{\nu})  \approx 1,
\end{equation}
where $P_{\nu}$ is the projection operator onto $\mathscr{H}_{\nu}$. This means that almost all of $W$ is in $\mathscr{H}_{\nu}$. 
\item[7'] Thermal equilibrium: typically, there is a dominant macro-space $\mathscr{H}_{eq}$ that has a dimension that  is almost equal to D: 
\begin{equation}
\frac{\text{dim} \mathscr{H}_{eq}}{\text{dim} \mathscr{H}} \approx 1.
\end{equation}
A system with density matrix $W$ is in equilibrium if  $W $ is very close to $\mathscr{H}_{eq}$ in the sense of (\ref{Wclose}):  $\text{tr} (W P_{eq})  \approx 1$.
\item[8'] Boltzmann Entropy: the Boltzmann entropy of a quantum-mechanical system with density matrix $W$ that is very close to a macrostate $\nu$ is given by:
\begin{equation}\label{Boltzmann}
S_B (W) = k_B \text{log} (\text{dim} \mathscr{H}_\nu ),
\end{equation}
where $\mathscr{H}_\nu$ denotes the subspace containing almost all of $W$ in the sense of (\ref{Wclose}). 
\end{itemize}

For the Bohmian version, we have the additional resource of the particle configuration, which can enable us to further  define the effective macrostate and the effective Boltzmann entropy of a quantum system in a way that is analogous to the effective wave function of the system.\footnote{Here is a flat-footed way to do it. 
First, we start from the wave-function picture of BM. If $\Psi_i$ is the effective wave function of the universe, and if $\Psi_i$ is almost entirely in $\mathscr{H}_{\nu}$, then the universe is in an effective macrostate of $\mathscr{H}_{\nu}$ and it has effective Boltzmann entropy of $k_B \text{log} (\text{dim} \mathscr{H}_\nu )$. 
Second, we note that we can define an analogous notion of the effective density matrix of the universe. 
Third, we can transpose the first point to the density-matrix perspective. 
If $W_i$ is the effective density matrix of the universe, and if $W_i$ is almost entirely in $\mathscr{H}_{\nu}$, then the universe is in an effective macrostate of $\mathscr{H}_{\nu}$ and it has effective Boltzmann entropy of $k_B \text{log} (\text{dim} \mathscr{H}_\nu )$. 
To be sure, further analysis is required. }

\subsection{Conditional Uniqueness}

\subsubsection{Eliminating Statistical Postulate with Conditional Uniqueness}

The Past Hypothesis,  given a choice of an initial subspace, is compatible with a continuous infinity of initial wave functions. However, the Initial Projection Hypothesis determines a unique initial density matrix given an initial subspace; the initial quantum state is the normalized projection onto the subspace we select. I call this feature the \emph{conditional uniqueness} of the initial quantum state in $W_{IPH}$ theories. 

In the standard theory, the Statistical Postulate determines a uniform probability distribution ($\mu_S$ with a normalization constant) over wave functions once we fix an initial subspace. This is because there are some wave functions compatible with PH that are anti-entropic, i.e. they will evolve into lower-entropy states. Having a uniform probability distribution will provide a way to say that they are unlikely, since there are very few of them compared to the entropic wave functions. In the new theory ($W_{IPH}$), if we fix an initial subspace, there is only one choice for the initial quantum state because of conditional uniqueness.  If typical (in the sense of $\mu_S$) wave functions starting in  $\mathscr{H}_{PH}$ will increase in entropy, then we know that almost all of the normalized projection  onto $\mathscr{H}_{PH}$ will increase in entropy. 

Therefore, in $W_{IPH}$ theories, we no longer need the Statistical Postulate;  we no longer need a probability / typicality distribution over initial quantum states, since only one state is possible (given an initial subspace). The only kind of probability in those theories will be the quantum mechanical probabilities. We expect the following to be true. In $W_{IPH}$-BM, by the W-version of the quantum equilibrium postulate (see Appendix), most initial particle configurations will increase in entropy. In $W_{IPH}$-GRW, by the W-version of the spontaneous collapse postulate, most likely the future collapses will result in higher-entropy states. In $W_{IPH}$-EQM, most emergent branches / worlds will evolve into higher-entropy ones.\footnote{Much more work need to be done to show these results rigorously. A step in this direction is to show the empirical equivalence between $\Psi_{PH}$ theories and $W_{IPH}$ theories. For an overview of some existing results and some new analysis,  see \cite{chen2019quantum1}.}

The property ``conditional uniqueness" is a conditional property: the uniqueness of the initial quantum state depends on a choice of an initial subspace, representing the PH low-entropy macrostate. There are reasons to believe that the choice of the initial subspace will not be unique in the standard wave-function theory $\Psi_{PH}$. After all, the PH selects a low-entropy macrostate given macroscopic variables such as volume, density, and volume (\S2.2 \#5). The values of the macroscopic variables do not  have precise correspondence with subspaces in the Hilbert space. Two slightly different subspaces $\mathscr{H}_1$ and $\mathscr{H}_2$ that share most wave functions but differ regarding some wave functions will not make a difference to the role that PH plays. The role of PH is to explain the macroscopic regularity of entropic arrow of time and perhaps other arrows of time, which by themselves do not require a maximal level of precision. (Macroscopic facts are notoriously vague and imprecise. See the sorites problem  surveyed in \cite{sep-sorites-paradox}.) The precision in the subspace will not correspond to anything in the physical world. If the actual wave function $\Psi$ is contained in two slightly different subspaces $\mathscr{H}_1$ and $\mathscr{H}_2$, the actual velocity of Bohmian particles, the actual collapse probability distribution, and the structure of emergent Everettian branches  are all determined by $\Psi$. The slight differences between the two subspaces (i.e. slight differences in the boundary of the macrostate) have no influence on the microscopic facts in the world, since any microscopic influence is screened off by the actual wave function. Moreover, the non-unique correspondence in QSM (\S2.2 \$6) makes the boundaries between macrostates even more fuzzy. It would seem highly arbitrary to pick a precise boundary of the macrostate. If there are many choices of the initial subspace, all differing from each other by a small amount, then perhaps there will be many choices of the initial projection operators that are hard to distinguish on a macroscopic level.  

However, if there are merely multiple choices of the initial subspace, and it is \emph{determinate} which ones are the admissible choices,  perhaps we can just collect all the admissible choices of the initial subspace and make PH a disjunctive statement: it is one of the admissible choices. But the problem is that what is admissible and what is not admissible will also be \emph{indeterminate. } This is analogous to the higher-order problem in vagueness (see \cite{sep-vagueness}).  So the most realistic version of PH will be allowing the  fuzzy boundary of the initial macrostate. We discuss these different versions in more detail below.

\subsubsection{Three Versions of PH and IPH}

In what follows, I formulate three versions of PH, with different levels of precision: the Strong PH, the Weak PH, and the Fuzzy PH. For each version, I formulate  a version of IPH that matches the level of precision in PH. Although the Fuzzy PH might be the best choice for the standard theory, the Strong IPH might be viable for the new theories of $W_{IPH}$. 


The Strong Past Hypothesis relies on a particular decomposition of the energy shell into macrostates (\S2.2 \#5). It selects a unique initial subspace---$\mathscr{H}_{PH}$. The initial wave function has to start from $\mathscr{H}_{PH}$. 
\begin{description}
\item[Strong Past Hypothesis] 
\end{description}
\begin{equation} 
\Psi(t_0) \in \mathscr{H}_{PH}
\end{equation}
Moreover, it is a well-defined probability space on which we can impose the surface area measure $\mu_S$ (on the unit sphere of vectors). If the Strong PH makes into the $\Psi_{PH}$ theories, then we can impose a similarly strong IPH in the $W_{IPH}$ theories:
\begin{description}
\item[Strong Initial Projection Hypothesis] 
\end{description}
\begin{equation}
\hat{W}_{IPH} (t_0)  = \frac{I_{PH}}{dim \mathscr{H}_{PH}}
\end{equation}
The Strong IPH selects a unique initial quantum state, which does not require a further specification of statistical mechanical probabilities. 

As mentioned before,  there may be several low-dimensional subspaces that can do the job of low-entropy initial condition equally well. They do not have to be orthogonal to each other. The Past Hypothesis is formulated in macroscopic language which is not maximally precise. The inherent imprecision can result in indeterminacy of the initial subspace. There are two kinds of indeterminacy here: (1) there is a determinate set of admissible subspaces to choose from, and (2) the boundary between what is admissible and what is not is also fuzzy; some subspaces  are admissible but some others are borderline cases. 

The first kind of indeterminacy corresponds to the Weak Past Hypothesis:
\begin{description}
\item[Weak Past Hypothesis] 
\end{description}
\begin{equation} 
\Psi(t_0) \in \mathscr{H}_{1} \text{ or } \Psi(t_0) \in \mathscr{H}_{2} \text{ or ... or }  \Psi(t_0) \in \mathscr{H}_{k}
\end{equation}
To be sure, the set of admissible subspaces may be infinite. For simplicity we shall assume it is finite for now. For the Weak PH, there is the corresponding  Weak IPH:
\begin{description}
\item[Weak Initial Projection Hypothesis] 
\end{description}
\begin{equation}
\hat{W}_{WPH_i} (t_0)  = \frac{I_{\mathscr{H}_{i}}}{dim \mathscr{H}_{i}}
\end{equation}
where $ \mathscr{H}_{WPH_i}$ is the normalized projection onto the subspace $\mathscr{H}_i$. For each admissible initial subspace, there is exactly one choice of the initial quantum state. Even though there is only one actual initial density matrix, there are many  possible ones that slightly differ from each other ($\hat{W}_{WPH_1} (t_0)$, $\hat{W}_{WPH_2} (t_0)$, ..., $\hat{W}_{WPH_k} (t_0) $). Weak IPH does not need to be supplemented with an additional probability distribution over possible initial quantum states in that subspace. 

The second kind of indeterminacy corresponds to the Fuzzy Past Hypothesis:
\begin{description}
\item[Fuzzy Past Hypothesis] 
\end{description}
The universal wave function started in some low-entropy state. The boundary of the macrostate is vague. (The Fuzzy PH does not determinately pick out any set of subspaces.)

Given the Fuzzy PH, we can formulate a Fuzzy IPH:
\begin{description}
\item[Fuzzy Initial Projection Hypothesis] 
\end{description}
\begin{equation}
\hat{W}_{FPH} (t_0)  = \frac{I_{\mathscr{H}_{i}}}{dim \mathscr{H}_{i}}
\end{equation}
where $\mathscr{H}_{i}$ is an admissible precisification of the Fuzzy PH. Just like the Weak IPH, the Fuzzy IPH does not select a unique initial quantum state, since there are many admissible ones. (There is an actual initial density matrix, and it is compatible with the Fuzzy IPH.) However, unlike Weak IPH, there is no sharp cut-off between what is admissible and what is not admissible. Hence, for theories that include the Fuzzy IPH, it may be indeterminate whether some normalized projection operator is possible or not possible. 

In any case, neither the Weak IPH nor the Fuzzy IPH   requires the further specification of statistical mechanical probabilities. To see why,  consider what is required for the Statistical Postulate. It requires a probability space to define the measure. The measure can only be defined relative to an admissible precisification. That is, the measure is over possible state vectors in a precise state space. In other words, the statistical mechanical probabilities are only meaningful relative to a precisification of the initial subspace. To be sure, the exact probability may be empirically inaccessible to us --- what we have access to may only be certain interval-valued probabilities.\footnote{See \cite{sep-imprecise-probabilities} for a  review of imprecise probabilities.} However, without the precisification of the boundary, we do not even have an interval-valued probabilities, as where to draw the boundary is unsettled. Given this fact, we can see that the original statistical mechanical probabilities require admissible precisifications---the statement will have to be:
\begin{itemize}
\item For a choice of an initial subspace, and given a uniform measure $\mu_S$ on it,  $\mu_S-$most (maybe even all) wave functions in that subspace will approach thermal equilibrium. 
\end{itemize}
So the corresponding statement for the Weak IPH and the Fuzzy IPH will be:
\begin{itemize}
\item Every admissible density matrix will approach thermal equilibrium, where an admissible density matrix is a normalized projection operator onto an admissible initial subspace.
\end{itemize}
Since every admissible density matrix will approach thermal equilibrium, there is no need for an additional probability measure over initial quantum states in order to neglect the anti-entropic states. What is different on Fuzzy IPH is that what is admissible is a vague matter.

To summarize: all three versions of the IPH get rid of the fundamental statistical mechanical probabilities. The Strong IPH selects a unique initial quantum state. Although the Weak IPH and the Fuzzy IPH do not, the initial state is  unique given an admissible precisification of the PH subspace. Hence, all three versions---Strong IPH, Weak IPH, and Fuzzy IPH---satisfy conditional uniqueness of the initial quantum state. That is, given a choice of an initial subspace, there is a unique initial quantum state, i.e. its normalized projection. Since conditional uniqueness is sufficient to eliminate statistical mechanical probabilities at the fundamental level, all three versions will be able to reduce the two kinds of probabilities to just the quantum mechanical probabilities. However, for Strong IPH, the choice of the initial subspace is unique while it is not unique for Weak IPH or Fuzzy IPH. Therefore, Strong IPH  trivially satisfies conditional uniqueness.

\subsubsection{The Emergence of Statistical Mechanical Probabilities}

Although the $W_{IPH}$ theories (with any of the three versions of IPH) eliminate statistical mechanical probabilities at the fundamental level, they can still emerge as useful tools for analysis. As we saw in (\ref{MacroW}), the normalized projection onto a subspace can be decomposed as an integral of pure state density matrices with respect to the uniform probability distribution on the unit sphere. So mathematical results proven for a statistical ensemble of wave functions can be directly applied to normalized projections. The statistical perspective could prove useful for analysis of typicality and long-time behaviors of a statistical ensemble of wave functions, which could be translated to the long-time behavior of the mixed-state density matrix. 

However, we recall that the decomposition into an integral of pure states is not unique. For the same density matrix in  (\ref{MacroW}), it can be decomposed as a sum of pure states that are the orthonormal basis vectors, which gives us a different statistical ensemble of wave functions (a discrete probability distribution). Different statistical ensembles and the different emergent probabilities can be useful for different purposes. The non-uniqueness in decomposition further illustrates the idea that, in $W_{IPH}$ theories, given an objective (real) density matrix $W$, the statistical mechanical probabilities over initial wave functions are not fundamental but are only emergent at the level of analysis.

\subsection{Absolute Uniqueness}

All three versions of IPH satisfy the property called \emph{conditional uniqueness}: given a choice of the initial subspace, there is a unique choice of the initial quantum state---the normalized projection onto that subspace. The Strong IPH satisfies that trivially. Indeed, the Strong IPH satisfies a stronger property that I call \emph{absolute uniqueness}: there is only a unique choice of the initial quantum state. This is because Strong IPH builds in a selection of a unique initial subspace. 

As we mentioned in \S3.2.1, the Strong PH in the $\Psi_{PH}$ theories, which is the counterpart to the Strong IPH in $W_{IPH}$ theories, is highly implausible. The precision of the Strong PH corresponds to a precise low-entropy initial macrostate, which has no correspondence to any precision in the microscopic histories of particles, collapses, or field values. The PH macrostate plays merely a macroscopic role, and the actual microstate, i.e. a wave function,  screens off the microscopic influences of the PH macrostate. 

However, the situation is quite different in $W_{IPH}$ theories with a Strong IPH. Here, the initial density matrix $W_{IPH} (t_0)$ plays two roles. First, it plays the usual  macroscopic role corresponding to the low-entropy initial condition. Second, and more importantly, it plays the microscopic role since it enters into the fundamental dynamical equations of $W_{IPH}$ quantum mechanics (see Appendix for details). $W_{IPH} (t_0)$ directly guides Bohmian particles, undergoes GRW collapses, and realizes the emergent branches of Everett. Different choices of $W_{IPH} (t_0)$ will leave different microscopic traces in the world, in the form of different velocity fields for Bohmian particles, different probabilities for GRW collapses, and emergent branches with microscopic differences. Hence, the extra precision that sets Strong IPH apart from Weak or Fuzzy IPH has a worldly correspondence: it matches the microscopic  history of the universe. Even if we were to think that different choices of the initial subspaces might not make an observable differences (at least to our current observational technologies), there is still a fact of the matter in the world about the precise microscopic  history. Hence,  it seems to me  it is quite justifiable to postulate the Strong IPH in $W_{IPH}$ theories, even though it is not justifiable to postulate the Strong PH in $\Psi_{PH}$ theories.

The Strong IPH has another interesting consequence. It will make the Everettian theory strongly deterministic in the sense of \cite{roger1989emperor}: Strong determinism
 ``is not just a matter of the future being determined by the past; \emph{the entire history of the universe is fixed}, according to some precise mathematical scheme, \emph{for all time.}''  This is because the Everettian theory is deterministic for all beables (local or non-local). With the help of a unique initial state, there is no ambiguity at all about what the history could be---there is only one way for it to go: starting from $\hat{W}_{IPH} (t_0)$ (on Strong IPH) and evolving by the von Neumann equation (\ref{VNM}).

\subsection{Other Applications}

Getting rid of statistical mechanical probabilities at the fundamental level is not the only advantage of $W_{IPH}$-quantum theories over $\Psi_{PH}$-quantum theories. I discuss them in more detail in \cite{chen2018IPH, chen2018HU}. Here I briefly summarize the other features of $W_{IPH}$-quantum theories: 
\begin{enumerate}
\item The meaning of the quantum state.  

It has been a long-standing puzzle regarding how to understand the meaning of the quantum state, especially because it is defined as a function on a high-dimensional space and it is a non-separable object in physical space. A particularly attractive proposal is to understand it as nomological---being on a par with laws of nature. However, it faces a significant problem since typical wave functions are highly complex and not simple enough to be nomological. In contrast, if the quantum state is given by IPH, the initial quantum state will inherit the simplicity of the PH subspace. If PH is simple enough to be a law, then the initial quantum state is simple enough to be nomological. This feature of simplicity supports  the nomological interpretation  without relying on specific proposals about quantum gravity (cf: \cite{goldstein2013reality}). However, I should emphasize that the nomological interpretation is not the only way to understand the density matrix theory, as other proposals are also valid, such as the high-dimensional field and the low-dimensional multi-field interpretations (\cite{chen2018IPH} \S3.3). 

\item Lorentz invariance. 

David \cite{albert2015after}  observes that there is a  conflict among three properties: quantum entanglement, Lorentz invariance, and what he calls narratability. A world is narratable just in case its entire history can be narrated in a linear sequence and every other sequence is merely a geometrical transformation from that. Since narratability is highly plausible, denying it carries significant cost. Hence, the real conflict is between quantum entanglement (a purely kinematic notion) and Lorentz invariance. This applies to all quantum theories that take quantum entanglement to be fundamental. However, given IPH, it is possible to take the nomological interpretation of the quantum state and remove entanglement facts among the facts about the distribution of physical matter. This is especially natural for $W_{IPH}$-Sm,  somewhat less naturally to $W_{IPH}$-GRWm and $W_{IPH}$-GRWf, and potentially applicable to a fully Lorentz-invariant version of $W_{IPH}$-BM. 

\item Kinematic and dynamic unity. 

In $\Psi_{PH}$-quantum theories, many subsystems will not have pure-state descriptions by wave functions due to the prevalence of entanglement. Most subsystems can be described only by a mixed-state density matrix, even when the universe as a whole is described by a wave function. In contrast, in W$_{IPH}$-quantum theories, there is more uniformity across the subsystem level and the universal level: the universe as a whole as well as most subsystems are described by the same kind of object---a (mixed-state) density matrix. Since state descriptions concern the kinematics of a theory, we say that W$_{IPH}$-quantum theories have more \emph{kinematic unity} than their $\Psi$-counterparts:
\begin{description}
\item \textsc{Kinematic Unity} The state description of the universe is of the same kind as the state descriptions of most quasi-isolated subsystems. 
\end{description} 
Moreover, in a universe described by $\Psi_{PH}$-BM,  subsystems sometimes do not have conditional wave functions due to the presence of spin. In contrast, in a universe described by $W_{IPH}$-BM, the universe and the subsystems always have quantum states given by density matrices. This is because we can always define conditional density matrix for a Bohmian subsystem (\cite{durr2005role}). That is, in $W_{IPH}$-BM, the W-guidance equation is always valid for the universe and the subsystems. In $\Psi_{PH}$-BM, the wave-function version of the guidance equation is not always valid. Thus, the W-BM equations are valid in more circumstances than the BM equations. We say that $W_{IPH}$-BM has more dynamic unity than $\Psi_{PH}$-BM: 
\begin{description}
\item \textsc{Dynamic Unity} The dynamical laws of the universe are the same as the effective laws of most quasi-isolated subsystems. 
\end{description} 
Both kinematic unity and dynamic unity come in degrees, and they are only defeasible reasons to favor one theory over another. But it is nonetheless interesting that merely adopting the density-matrix framework will make the theory more ``unified'' in the above senses. 

\end{enumerate}

\subsection{Generalizations}

IPH is not the only principle that leads to the selection of a unique or effectively unique initial quantum state of the universe. It is just one example of a simple principle. It is easy to generalize the strategy discussed here to other simple principles about the cosmological initial condition:
\begin{itemize}
\item Start from the full Hilbert space (energy shell) $\mathscr{H}$. 
\item Use simple principles (if there are any) to determine an initial subspace $\mathscr{H}_0 \subset \mathscr{H}$. 
\item Choose the natural quantum state in that subspace---the normalized projection $\hat{W}_{0} (t_0) = \frac{I_{0}}{dim \mathscr{H}_{0}}$. 
\item The natural choice will be simple and unique.
\end{itemize}

Another related but different example is the quantum version of the Weyl Curvature Hypothesis proposed by \cite{ashtekar2016initial} based on the proposal of the classical version in \cite{roger1989emperor}. However, Ashtekar and Gupt's hypothesis results in an infinite-dimensional unit ball of initial wave functions, which may not be normalizable. It is not clear to me whether there will be significant dimension reduction if we intersect it with the energy shell. In any case, the problem of non-normalizability is a general problem in cosmology which may require an independent solution.

\subsection{Other Proposals}

I would like to contrast my proposal with three other proposals in the literature. 

Albert (2000) proposes that it is plausible that  $\Psi_{PH}$-GRW theories do not need the Statistical Postulate. This is because the GRW jumps may be large enough (in Hilbert space) to render every initial wave function entropic in a short time. An anti-entropic wave function of a macroscopic system that evolves forward will be quickly hit by a GRW jump. As long as the GRW jump has a certain width that is  large compared to  the width of the anti-entropic set, the wave function will collapse into an entropic one. This relies on a conjecture about GRW theory: the final and empirically adequate GRW models will have a collapse width that is large enough. If that conjecture can be established, then, for every initial wave function, it is with a GRW probability that it will be entropic. This may correspond to the right form of the probabilistic version of the Second Law of Thermodynamics. I believe this is a very plausible conjecture. But it is an additional conjecture nonetheless, and it only works for collapse theories. It is an empirically open question whether GRW will survive experimental tests in the next 30 years. In contrast, my proposal is fully general---it works for GRW theories, Bohmian theories, and Everettian theories, and it does not rely on additional conjectures beyond those already postulated in QSM. 

\cite{wallace2011logic} proposes that we can replace the Past Hypothesis and the Statistical Postulate with a Simple Dynamical Conjecture. In essence, it says that every Simple wave function will evolve to higher entropy. Here, ``Simple'' is a technical notion here meaning that the wave function is simple to specify and not specified by using time-reverse of an anti-entropic wave function. The idea is to replace the statistical postulate, which gives us reasons to neglect certain initial wave functions, with another postulate about simplicity, which also gives us reasons to set aside certain initial wave functions. This is a very interesting conjecture, which I think one should seriously investigate. But it is an additional conjecture nonetheless, although it has applicability to all quantum theories. 

\cite{wallace2016probability} proposes that we can allow quantum states to be either pure or mixed. Moreover, he suggests that we can  reinterpret probability distributions over wave functions as part of the state description and not an additional postulate about probability. There is much in common between Wallace's (2016) proposal and my proposal.  However, the way that I get rid of statistical mechanical probabilities is not by way of a reinterpretation strategy but by using the uniqueness (or conditional uniqueness) of the initial quantum state. To be sure, there are many ways to achieve the goal, and the two approaches are quite related. 

I would like to briefly mention the possibility of implementing my proposal  in CSM. The upshot is that it is much less natural to do so in CSM than in QSM, for several reasons. On the classical mechanical phase space, the object that plays a similar role to the density matrix is the probability function $\rho(x)$. However, it is not clear how to understand its meaning as something ontic. If it is to be understood as a high-dimensional field and the only object in the ontology, then we have a Many-Worlds theory, unless we revise the Hamiltonian dynamics and make it stochastic. If it is to be understood as a low-dimensional multi-field (cf: \cite{ChenOurFund} and \cite{Hubert2018} and the references therein), it is not clear what corresponds to the momentum degrees of freedom. 
Second, the dynamics is not as natural in CSM. If it is to be understood as a high-dimensional field guiding a point particle (the phase point), it is not clear why we increase the complexity of the ontology. If it is to be understood as a nomological object, it is not clear what role it plays in the dynamics---it certainly does not give rise to a velocity field, because that is the job of the Hamiltonian function.\footnote{See \cite{McCoySMS}  for a recent attempt in this direction. Despite what I say here, I believe it is worth exploring, as McCoy does, alternative interpretations of the statistical state in CSM and QSM. }

\section{Conclusion}

The Thermodynamic Theories of Quantum Mechanics (TQM or $W_{IPH}$-QM) provide a new strategy to eliminate statistical mechanical probabilities from the fundamental postulates of the physical theory. They do so in a simple way not relying on any further conjectures about quantum mechanics or statistical mechanics. Moreover, they lead to several other applications and generalizations, which we only touched on briefly in this paper. Most importantly, we have found another deep connection between the foundations of quantum mechanics and the foundations of statistical mechanics. 
 
\section*{Appendix}

\subsection*{(1) $\Psi_{PH}$-Quantum Theories}

\subsubsection*{$\Psi_{PH}$-Bohmian mechanics: ($Q, \Psi_{PH}$)}

The Past Hypothesis:
\begin{equation}
\Psi_{PH} (t_0)  \in  \mathscr{H}_{PH}
\end{equation}

The Initial Particle Distribution: 
\begin{equation}\label{QEH}
\rho_{t_0} (q) = |\psi(q, t_0)|^2
\end{equation}

The Schr\"odinger Equation:
\begin{equation}\
 i\hbar \frac{\partial \psi}{\partial t} = H \psi
\end{equation}

The Guidance Equation (D{\"u}rr et al. 1992):
\begin{equation}\label{GE}
 \frac{dQ_i}{dt} = \frac{\hbar}{m_i} \text{Im} \frac{ \nabla_i \psi (q) }{  \psi (q)} (q=Q)
\end{equation}

\subsubsection*{$\Psi_{PH}$-Everettian mechanics}

The Past Hypothesis:
\begin{equation}
\Psi_{PH} (t_0)  \in  \mathscr{H}_{PH}
\end{equation}

The Schr\"odinger Equation:
\begin{equation}
 i\hbar \frac{\partial \psi}{\partial t} = H \psi
\end{equation}

The Mass Density Equation: 
\begin{equation}\label{mxt}
m(x,t) = \bra{\psi(t)} M(x) \ket{\psi(t)},
\end{equation}

$W_{PH}$-S0: only $\Psi_{PH}$. 

$W_{PH}$-Sm: $m(x,t)$ and $\Psi_{PH}$. 

\subsubsection*{$\Psi_{PH}$-GRW theory}

The Past Hypothesis:
\begin{equation}
\Psi_{PH} (t_0)  \in  \mathscr{H}_{PH}
\end{equation}

The linear evolution of the wave function is interrupted randomly (with rate $N\lambda$, where $\lambda$ is of order $10^{-15}$ s$^{-1}$) by collapses: 
\begin{equation}\label{WFcollapse}
\Psi_{T^+} = \frac{\Lambda_{I_{k}} (X)^{1/2} \Psi_{T^-} }{||  \Lambda_{I_{k}} (X)^{1/2} \Psi_{T^-}  || },
\end{equation}
with the collapse center $X$ being chosen randomly with probability distribution $\rho(x) = ||  \Lambda_{i_{k}} (x)^{1/2} \Psi_{T^-}  ||^2 dx$, 
where the collapse rate operator is defined as:
\begin{equation}
\Lambda_{I_{k}} (x) = \frac{1}{(2\pi \sigma^2)^{3/2}} e^{-\frac{(Q_k -x)^2}{2\sigma^2}}
\end{equation}
where $Q_k$ is the position operator of ``particle'' $k$, and $\sigma$ is a new constant of nature of order $10^{-7}$ m postulated in current GRW theories. 

$\Psi_{PH}$-GRWm and $\Psi_{PH}$-GRWf: defined with local beables $m(x,t)$ and $F$.

\subsection*{(2) $W_{IPH}$-Quantum Theories}

\subsubsection*{$W_{IPH}$-Bohmian mechanics: ($Q, W_{IPH}$)}

The Initial Projection Hypothesis:
\begin{equation}
\hat{W}_{IPH} (t_0)  = \frac{I_{PH}}{dim \mathscr{H}_{PH}}
\end{equation}

The Initial Particle Distribution: 
\begin{equation}
P(Q(t_0) \in dq) =  W_{IPH} (q, q, t_0) dq
\end{equation}

The Von Neumann Equation:
\begin{equation}
i \hbar \frac{\partial \hat{W}}{\partial t} = [\hat{H},  \hat{W}]
\end{equation}

The $W_{PH}$-Guidance Equation (D{\"u}rr et al. 2005):
\begin{equation}
\frac{dQ_i}{dt} = \frac{\hbar}{m_i} \text{Im} \frac{\nabla_{q_{i}}  W_{IPH} (q, q', t)}{ W_{IPH} (q, q', t)} (q=q'=Q)
\end{equation}

\subsubsection*{$W_{IPH}$-Everettian mechanics}
The Initial Projection Hypothesis:
\begin{equation}
\hat{W}_{IPH} (t_0)  = \frac{I_{PH}}{dim \mathscr{H}_{PH}}
\end{equation}

 The Von Neumann Equation:
\begin{equation}
i \hbar \frac{\partial \hat{W}}{\partial t} = [\hat{H},  \hat{W}]
\end{equation}

The Mass Density Equation: 
\begin{equation}\label{mxt}
m(x,t) = \text{tr} (M(x) W(t)),
\end{equation}

$W_{IPH}$-S0: only $W_{IPH}$. 

$W_{IPH}$-Sm: $m(x,t)$ and $W_{IPH}$. 

\subsubsection*{$W_{IPH}$-GRW theory}

The Initial Projection Hypothesis:
\begin{equation}
\hat{W}_{IPH} (t_0)  = \frac{I_{PH}}{dim \mathscr{H}_{PH}}
\end{equation}

The linear evolution of the density matrix is interrupted randomly (with rate $N\lambda$, where $\lambda$ is of order $10^{-15}$ s$^{-1}$) by collapses: 
\begin{equation}\label{collapse}
W_{T^+} = \frac{\Lambda_{I_{k}} (X)^{1/2} W_{T^-} \Lambda_{I_{k}} (X)^{1/2}}{\text{tr} (W_{T^-} \Lambda_{I_{k}} (X)) }
\end{equation}
with $X$ distributed by the following probability density:
\begin{equation}\label{center}
\rho(x) = \text{tr} (W_{T^-} \Lambda_{I_{k}} (x))
\end{equation}
where the collapse rate operator is defined as:
\begin{equation}
\Lambda_{I_{k}} (x) = \frac{1}{(2\pi \sigma^2)^{3/2}} e^{-\frac{(Q_k -x)^2}{2\sigma^2}}
\end{equation}
where $Q_k$ is the position operator of ``particle'' $k$, and $\sigma$ is a new constant of nature of order $10^{-7}$ m postulated in current GRW theories. The dynamical equations of W-GRW were introduced in \cite{allori2013predictions} \S4.5.

$W_{IPH}$-GRWm and $W_{IPH}$-GRWf: defined with local beables $m(x,t)$ and $F$.


\section*{Acknowledgement}

I would like to thank two anonymous reviewers of this volume for their helpful feedback on an earlier draft of this paper. I am also grateful for stimulating discussions with Valia Allori, Sean Carroll, Detlef D\"urr,  Michael Esfeld,  Veronica Gomez, Dustin Lazarovici, Matthias Lienert, Tim Maudlin, Sebastian Murgueitio, Wayne Myrvold, Jill North, Daniel Rubio, Ted Sider,  Roderich Tumulka, David Wallace, Isaac Wilhelm, Nino Zangh\`i, and especially David Albert, Sheldon Goldstein, and Barry Loewer.

\bibliography{test}


\end{document}